\documentclass[a4paper]{jpconf}
\usepackage{amsfonts}
\usepackage{amsmath}

\input{tcilatex}
\font\af=msbm11

\def\R{\hbox{\af R}}
\def\I{\hbox{\af I }}
\def\be{\begin{equation}}
\def\ee{\end{equation}}

\begin{document}

\title{Boundary conditions: The path integral approach}
\author{M. Asorey $^1$,   J. Clemente-Gallardo $^2$ and J.M. Mu\~noz-Casta\~neda$^1$ }
\address{$^1$ 
Departamento de F\'{\i}sica Te\'{o}rica, Universidad de Zaragoza
50009 Zaragoza, Spain }
\address{$^2$ 
 BIFI,
Universidad de Zaragoza, 
50009 Zaragoza, Spain }
\ead{%
asorey@unizar.es, jcg@unizar.es, jmmc@unizar.es}

\begin{abstract}
The path integral approach to quantum mechanics requires a 
 substantial generalisation  to describe the dynamics of 
systems  confined to bounded domains. Non-local boundary conditions
can be introduced in  Feynman's approach by means of  boundary amplitude distributions
 and  complex phases  to describe the quantum dynamics in terms of the classical trajectories. 
The different prescriptions involve only  trajectories reaching the boundary and  correspond to different choices 
of boundary conditions of  selfadjoint extensions of the Hamiltonian. 
One dimensional particle dynamics is 
analysed  in  detail.
\end{abstract}

\section{Introduction}
\label{sec:1}
\medskip

The original approach to quantum mechanics due to Heisenberg and  Schr\"odinger is 
based on the Hamiltonian formalism. Dirac and later Feynman introduced a Lagrangian
approach. The  Feynman's path integral method was shown later to be equivalent to
the traditional formulation for most of quantum mechanical systems \cite{Feynman, Feynman-Higgs}. 
In field theory the method presents  some advantages over  Hamiltonian 
quantisation. The Lagrange formalism  preserves relativistic  covariance 
which makes the Feynman method  very convenient
to achieve the  renormalization of field theories both in 
perturbative and  non-perturbative approaches.
 In the Euclidean version \cite{FeyKac} 
the functional integral formulation  is a crucial ingredient for
non-perturbative  approaches to field theory  and critical phenomena
\cite{wilson}.  The equivalence of the functional integral formulation with the
Hamiltonian pictures also holds for constrained systems  like   gauge theories
and string theories. 

However, for  systems constrained  to bounded domains  Feynman's approach 
becomes more intricate than the Hamiltonian operator approach and require a sophisticated
implementation for generic  types of boundary conditions
 compatible with  Hamiltonian approach
\cite{sant}. The
analysis of this problem is the main goal of this paper. 
On the other hand the physics of quantum systems constrained 
to bounded domains is becoming very relevant not only for applications in
condensed matter (quantum Hall effect, graphene, quantum wires, quantum dots, etc) 
but  also in fundamental physics  in fields like quantum gravity and  string theory.
In all these phenomena the role of boundary conditions is very important 
to describe a variety of new physical effects like  anomalies \cite{anm, aae, esteve}
topology change \cite{rbal} quantum holography \cite{bek,qg,hol}, quantum gravity 
and AdS/CFT correspondence \cite{mal}.
To some extent  the relevance of boundaries in the description
of fundamental physical phenomena.
has  promoted the role of  boundary phenomena
 from  academic and phenomelogical simplifications
of  complex physical  systems to a higher status connected with
very basic fundamental principles.  The application of the path integral approach to those problems
requires a radical generalisation of Feynman's formalism.

\section{ Boundary conditions in quantum mechanics}
\label{sec:2}

\medskip 

Unitarity  is the basic quantum principle 
which governs   the  dynamics 
of quantum systems confined in  bounded domains. 
The analytical implementation of this condition  in the  selfadjoint character   of the 
Hamiltonian operator encodes  all the quantum subtleties associated
to the unitarity principle and the dynamical behaviour at the boundary.
The selfadjointness condition of the Hamiltonian  imposes severe restrictions to the 
boundary conditions which are compatible with quantum mechanics.

The existence of a boundary generically enhances the genuine quantum aspects 
of the system. Well known   examples of this enhancement are  Young's two slits experiments
and the Aharonov-Bohm effect, which pointed out the relevance of  boundary 
conditions  in the quantum theory. Another examples of quantum physical
phenomena which are intimately related to boundary conditions are the Casimir effect
\cite{cas, vac} the role of edge states \cite{sr} and the quantisation of conductivity \cite{qcd, 9bis}
in the quantum Hall effect.

  The quantum role of 
boundary conditions is very important for the behaviour of low energy
levels. High energy levels are quite independent of boundary effects. Indeed, 
boundary effects play no role in the ultraviolet  regime, whereas they are
crucial for the infrared \cite{karpacz}.

Let us consider a point-like particle moving on a bounded domain $\Omega$
of $\R^n$ with  smooth  oriented boundary $\partial\Omega$. The  Hamiltonian 
of the system can be constructed from the scalar Laplacian 
\be
H=\frac12\Delta=-\frac12\sum_{i=1}^n \partial^i \partial_i .
\ee
This operator is symmetric on the domain of smooth wave functions with compact support on
$\Omega$. However, the restriction of the Laplacian operator  to  this domain   
is not selfadjoint, but  it
can be extended to   larger 
domains of wave functions where it becomes selfadjoint. 
The extension, however, not unique.

The  classification of selfadjoint extensions of the Hamiltonian
can be characterised in terms of unitary operators between
 defect subspaces in the classical theory due to von Neumann \cite{ds,krein}.
However, there is a more useful  characterisation of these selfadjoint
 extensions in
terms of constraint conditions on the boundary values of physical  wave functions
\cite{aim}.
 In this framework
the set of {  self-adjoint extensions of the Hamiltonian
 is in one-to-one correspondence with the group of unitary operators of the Hilbert 
 space $L^2(\partial\Omega)$  of wave  functions of the boundary $\partial\Omega$
which are square integrable with respect to  the  Riemaniann
measure induced from the Euclidean metric of $\R^n$.}

Thus,  any unitary operator  $U$ of the Hilbert space $L^2(\partial\Omega)$   defines a selfadjoint
 quantum Hamiltonian $\Delta^U$. And conversely, any selfadjoint extension of
$H$ is associated to one unitary operator $U$ of this type \cite{aim}.
The domain of the selfadjoint Hamiltonian governed by $U$ is defined
 by  the wave functions which satisfy the boundary condition
\begin{equation}
 \varphi - i\dot\varphi = U(\varphi + i\dot\varphi),
\label{bc}
\end{equation}
where
$\varphi=\psi|_{\partial\Omega}  $ is the boundary value of the wave function  $\psi$
and $\dot \varphi$ its oriented  normal  derivative at the boundary $\partial\Omega$.
Through  the above characterisation, the set of self-adjoint extensions of the
Hamiltonian  inherits  the group structure of the group of unitary operators.
For spaces  of dimension higher than one the group of boundary conditions  
is an infinite dimensional group. 

Since we shall focus on one-dimensional spaces two domains are of special interest:
the compact interval  $\Omega=[0,1]$ and the half line $\Omega=[0.\infty)$. Although
the second domain  is non-compact it will be useful to illustrate the main problems
due to  its simplicity. In the first case the set of boundary conditions is $U(2)$ whereas
for the half line it is $U(1)$.

\section{Boundary conditions in path integrals}
\label{sec:3}

\medskip

In  Feynman's  approach to quantum mechanics the dynamics is governed by an action principle 
similar to that  which governs the classical  dynamics.  Whereas the classical dynamics is given,
according to the variational action principle, by stationary
trajectories of the classical action,  the quantum dynamics 
is automatically implemented in the path integral formalism by the weight that
the classical action provides for particle trajectories. In the Euclidean time $T=-it$ formalism
the evolution propagator for the free particle  is given by the path integral
\be
\displaystyle 
K_{_T}(x,y)={\rm e}^{-TH}(x,y)= 
\int_{_{x(T)=y \atop  x(0)=x}}\!\!\!\!\!\!\!\!\!\!\! [\delta  x(t)] \,\, {\rm e}^{-\frac12\int_0^T \, \dot{x}(t)^2 dt}.
\label{path}
\ee

 However,  for particles evolving  in a bounded domain the variational problem  is
not uniquely defined in classical mechanics. In order to have a deterministic evolution
it becomes  necessary to specify the evolution of the particles
after reaching the  boundary. 
 A similar ambiguity problem arises in  quantum mechanics. 

The boundary imposes more severe constraints on the
classical dynamics than in the quantum one. This is due to the point-like
nature of the particle which requires that after reaching the boundary the
individual particle has to reemerge back either at  the same point   or at a  
different point of the boundary. In pure classical mechanics the only freedom concerns where it 
reemerges back  and the momentum it reemerges with. The emergence 
of the particle at a different point grants  the possibility 
of  folding and gluing  the boundary of the domain  giving rise to
non-trivial topologies. In summary, the classical boundary
conditions are given by two maps \cite{aim2}:  an isometry of the boundary
$\alpha:\partial\Omega\to\partial\Omega $
and a positive density function
$ \rho:\partial\Omega \to {\rm \R^+}$
which specify the change of position and normal component of 
momentum of the trajectory of the particle upon reaching the boundary. 
The isometry $\alpha$ encodes the possible geometry and
topology generated by the folding 
of the boundary and the function $\rho$  is associated to 
the reflectivity (transparency or stickiness) properties of the boundary. 
However, the quantum boundary conditions have 
a larger set of possibilities described by a unitary group.
In order to have a path integral description of all boundary
conditions we need to incorporate some random behaviour 
for the trajectories reaching the boundary  and complex phases
for those trajectories.  This is possible because the wave functions
are complex and the evolution operator involves complex amplitudes.
Although, is this way we are able to describe any type of
unitary evolution in the bounded domain the method goes  far beyond
Feynman's pure action approach.

The prescription is quite involved and proceeds by considering 
instead of the Euclidean time evolution propagator $K_{_T}$  the
resolvent operator $C_z$ of the Hamiltonian 
\be
\displaystyle C_z (x,y)={( z \I + H )^{-1}}(x,y)=\int_0^ \infty\frac{dT}{T} {\rm e}^{-zT} K_{_T}(x,y). 
\ee

The Euclidean time propagator can be recovered from the resolvent by means of the following countour integral
\be
K_{_T}(x,y)= \frac{1}{2\pi i}\oint \, C_z(x,y) {\rm e}^{zT} dz
\ee
along a contour which encloses the spectrum of $H$ on the real axis.

Boundary conditions can be  easily implemented into  the resolvent, whereas as we
shall see that the implementation in the Euclidean time propagator is much harder. 
Let us consider a fixed boundary condition, e.g.
the Neumann boundary conditions $U_0=\I$, and  consider the corresponding Hamiltonian $H^0$ as a
 background selfadjoint operator. The   selfadjoint extension of $H$ defined on the domain 
\be
 i(\I+U)\dot{\varphi}=(\I - U){\varphi}
\ee
by the unitary operator $U$
has a resolvent given by Krein's formula \cite{krein2}
\begin{equation}
\displaystyle C^U_z (x,y)= C^0_z (x,y) - \int_{\partial\Omega} dw\, \int_{\partial\Omega} dw'\,  C^0_z (x,w) R_z^U(w,w') C^0_z(w',y),
\label{krein2}
\end{equation}
 where $R^U$ is the operator of $ L^2(\partial\Omega) $ defined by 
\be
R_z^U=((\I - U) C^0_z -i (I+U))^{-1}(\I - U).
\ee

A similar formula could be obtained choosing another boundary condition as background
boundary condition instead of Neumann's condition.

The inverse transform permits to recover a formula for the propagator kernel of the type
\be
K_{_T}(x,y)= K^0_{_T}(x,y)- \frac{1}{2\pi i}\oint \, dz \,{\rm e}^{z T} \int_{\partial\Omega} dw\, \int_{\partial\Omega}
 dw'\,  C^0_z (x,w) R_z^U(w,w') C^0_z(w',y).
\ee
It is easy to rewrite  $K^0_{_T}(x,y)$ as a path integral as in  (\ref{path})
restricting the trajectories to the interior of the domain $\Omega$ and counting twice the trajectories hitting
the boundary  ${\partial\Omega}$. However, in general,  the kernel  $K_{_T}(x,y)$ cannot be rewritten
in terms of a path integral.  Only for a few boundary conditions the reduction can be achieved, but for generic
boundary conditions the kernel $K_{_T}(x,y)$ has to be considered as a genuine boundary condition kernel
containing information about the  boundary jumps amplitudes and phases associated to the different
trajectories hitting the boundary.  The complex structure of this kernel reduces the utility of the path integral
approach and points out the behaviour of the boundary as a genuine quantum device.
This behaviour can be explicitly
pointed out by noticing that under certain boundary conditions  the quantum evolution of a narrow
wave packet   is scattered  backward by  the boundary 
as a quite widespread wave packet emerging from all points of the boundary.

However, there are cases where this kernel adopts a simple form and the path integral approach
can be formulated in very explicit way. In particular, for  Dirichlet boundary conditions  $U=-\I$,
\be
R_z^U=(C^0_z )^{-1}
\ee
and 
\be
C^D_z(x,y) = C^0_z(x,y)  - \int_{\partial\Omega} dw\, \int_{\partial\Omega} dw'\,  C^0_z (x,w) (C^0_z)^{-1}(w,w') C^0_z(w',y),
\ee
which leads to a propagator  kernel given by the path integral  (\ref{path}) but restricted to paths which do not reach the
boundary  $\partial\Omega$.

Further  examples can be explicitly analysed for  one-dimensional domains.



\section{Particle in a half line}

\medskip

In order to illustrate the method  let us consider for simplicity the unbounded domain
$\Omega=[0,\infty]$. The set of selfadjoint extensions is parametrised by $U(1)$, i.e
the unitary operators of the form $U={\rm e}^{i\alpha}$. The corresponding boundary
conditions are the mixed conditions
\begin{equation}
\dot \varphi(0)=-\varphi'(0)= -\tan{\frac{\alpha}{2}}\varphi(0).
\end{equation}
The resolvent of the corresponding selfadjoint extension can be obtained from the
Krein formula (\ref{krein2}) giving rise to
\be
\displaystyle C^\alpha_z (x,y)=\frac{1}{2\sqrt{2 z}} {\rm e}^{-\sqrt{2z }|x-y|}+ \frac1{4 z} 
\frac{2z-\sqrt{2z}\, {\tan\frac{\alpha}{2}}} {\tan\frac{\alpha}{2}+\sqrt{2z}}
{\rm e}^{-\sqrt{2z }(|x|+|y|)},
\ee
which  in the two extreme cases $\alpha=0$ (Neumann)
\be
\displaystyle C^0_z (x,y)=\frac{1}{2\sqrt{2 z}} {\rm e}^{-\sqrt{2z }|x-y|}+ \frac1{2\sqrt{2z}} 
{\rm e}^{-\sqrt{2z }(|x|+|y|)}
\label{nnn}
\ee
and  $\alpha=\frac{\pi}{2}$ (Dirichlet)
\be
\displaystyle C^{\pi/2}_z (x,y)=\frac{1}{2\sqrt{2 z}} {\rm e}^{-\sqrt{2z }|x-y|}- \frac1{2\sqrt{2z}} 
{\rm e}^{-\sqrt{2z }(|x|+|y|)}
\label{ddd}
\ee
is reduced to very simple formulas.
Notice that because
\be
R^D_z= \sqrt{2z}
\ee
the Dirichlet resolvent $C_z^{\pi/2}$ can be easily obtained from the Neumann resolvent $C_z^{0}$ from Krein's formula 
(\ref{krein2}).

The inverse Laplace transform  gives the following formulas for the Euclidean evolution
kernels with Neumann boundary conditions \cite{sharp,pisani}
\be
\displaystyle K^N_{_T}(x,y)=\frac1{\sqrt{2\pi T}}{\rm e}^{-|x-y|^2/2T}+\frac1{\sqrt{2\pi T}}{\rm e}^{-|x+y|^2/2T}
\label{nn}
\ee
and Dirichlet boundary conditions
\be
\displaystyle K^D_{_T}(x,y)=\frac1{\sqrt{2\pi T}}{\rm e}^{-|x-y|^2/2T}- \frac1{\sqrt{2\pi T}}{\rm e}^{-|x+y|^2/2T},
\label{dd}
\ee
which can be interpreted in terms of the path integral as a sum over paths which never reach
the boundary \cite{sharp,pisani}, whereas the Neumann kernel corresponds to double-counting the paths which hit
the boundary
\footnote{In the Euclidean approach the restrictions imposed on the
paths for Neumann and Dirichlet boundary  conditions are interchanged 
with respect to the  boundary conditions in the Minkowski approach, where the
Dirichlet problem is associated to an infinite totally reflecting  wall.}. Notice that the terms depending on $x+y$,
which appear in  (\ref{nn}) (\ref{dd}) (\ref{nnn})and (\ref{ddd}), break
translation invariance due to the existence of finite  boundaries of the interval.

\section{Particle in an interval.}

\medskip

If the particle is confined in an interval $[0,1]$
the set of selfadjoint extensions is parametrised by $U(2)$. Although  the general theory
can be developed on the same basis, we shall consider only  few cases where the
path integral approach is simplified.

i) Neumann boundary conditions $U=\I$, $\dot\varphi=0$.

The resolvent with Neumann boundary conditions at both ends of the interval is (see e.g. \cite{albeverio,konstrykin}) 


\begin{equation}
\nonumber
\displaystyle C^{N}_z (x,y)= \frac{1}{2\sqrt{2 z}}\left[
\frac{{\rm e}^{-\sqrt{2z }(x+y)}}{1-{\rm e}^{-2\sqrt{2z }}}+
\frac{{\rm e}^{-\sqrt{2z }(2-x-y)}} {1-{\rm e}^{-2\sqrt{2z }}}.
+
 \frac{{\rm e}^{-\sqrt{2z }|x-y|}} {1-{\rm e}^{-2\sqrt{2z }}}
+\frac{{\rm e}^{-\sqrt{2z }(2-|x-y|)}} {1-{\rm e}^{-2\sqrt{2z }}} \right].
\end{equation}


The Euclidean time  propagator kernel 

\begin{eqnarray}
\nonumber
\displaystyle K^N_{_T}(x,y)= \frac{1}{\sqrt{2 \pi T}}\left[\sum_{n=-\infty}^\infty {\rm e}^{- (x-y+m)^2
/2T}+ \sum_{n=-\infty}^\infty {\rm e}^{- (x+y+m)^2/2T}\right]
\end{eqnarray}
corresponds to a path integral where the trajectories which hit the boundary are double weighted as
in the half line case.

ii) Dirichlet  boundary conditions $U=-\I$, $\varphi(0)=\varphi(1)=0$

The resolvent with Dirichlet boundary conditions at both ends of the interval is 


\begin{equation}
\nonumber
\displaystyle C^{D}_z (x,y)= \frac{1}{2\sqrt{2 z}}\left[
 \frac{{\rm e}^{-\sqrt{2z }|x-y|}} {1-{\rm e}^{-2\sqrt{2z }}}
+\frac{{\rm e}^{-\sqrt{2z }(2-|x-y|)}} {1-{\rm e}^{-2\sqrt{2z }}}
-
\frac{{\rm e}^{-\sqrt{2z }(x+y)}}{1-{\rm e}^{-2\sqrt{2z }}}-
\frac{{\rm e}^{-\sqrt{2z }(2-x-y)}} {1-{\rm e}^{-2\sqrt{2z }}}
\right].
\end{equation}


The associated  propagator kernel 

\begin{eqnarray}
\nonumber
\displaystyle K^D_{_T}(x,y)= \frac{1}{\sqrt{2 \pi T}}\left[\sum_{n=-\infty}^\infty {\rm e}^{- (x-y+m)^2
/2T}- \sum_{n=-\infty}^\infty {\rm e}^{- (x+y+m)^2/2T}\right]
\end{eqnarray}
%
corresponds to a path integral where the trajectories reaching the boundary have been removed.

iii) Periodic boundary conditions $U=\sigma_1=\begin{pmatrix}0&1\cr 1&0
\end{pmatrix}$, $\varphi(0)=\varphi(1)$, $\dot\varphi(0)=-\dot\varphi(1)$

The resolvent
\be
\displaystyle C^{p}_z (x,y)= \frac{1}{2\sqrt{2 z}}\left[ {\rm e}^{-\sqrt{2z }|x-y|}+ 
\frac{{\rm e}^{-\sqrt{2z }(1-x+y)}}{1-{\rm e}^{-\sqrt{2z}}}+
\frac{{\rm e}^{-\sqrt{2z }(1+x-y)}}{1-{\rm e}^{-\sqrt{2z}}}\right]
\ee
and the propagator kernel
\begin{eqnarray}
\nonumber
\displaystyle K^P_{_T}(x,y)= \frac{1}{\sqrt{2 \pi T}}\sum_{n=-\infty}^\infty {\rm e}^{- (x-y+n)^2
/2T}
\end{eqnarray}
indicate that the corresponding path integral is performed over periodic trajectories which cross from
one end of the interval to the other.

iv) Pseudo-periodic boundary conditions. 

The selfadjoint Hamiltonian corresponding to the unitary operator
\be
U=
\begin{pmatrix}
0&e^{-i\epsilon}\cr
e^{i\epsilon}&0
\end{pmatrix}
\ee
is defined in the domain satisfying  the boundary conditions
\be
{{\varphi(1)=e^{i\epsilon}\varphi(0)\qquad
\varphi'(1)=e^{i\epsilon}\varphi'(0)} }.
\ee

The resolvent 
\be
\displaystyle C^{\epsilon}_z (x,y)= \frac{{\rm e}^{i\epsilon(y-x)}}{2\sqrt{2 z}} 
\left[ {\rm e}^{-\sqrt{2z }|x-y|}+ \frac{{\rm e}^{-\sqrt{2z }(1-x+y)}}{1-{\rm e}^{-\sqrt{2z}-i\epsilon}}\, {\rm e}^{-i\epsilon}+
\frac{{\rm e}^{-\sqrt{2z }(1-y+x)}}{1-{\rm e}^{-\sqrt{2z}+i\epsilon}}\, {\rm e}^{i\epsilon}\right]
\ee
and  propagator 
\begin{eqnarray}
\nonumber
\displaystyle K^\epsilon_{_T}(x,y)= \frac{1}{\sqrt{2 \pi T}}\sum_{n=-\infty}^\infty {\rm e}^{- (x-y+n)^2/2T +i\epsilon (n+x-y)
}
\end{eqnarray}
kernels correspond to a path integral over periodic trajectories which are weighted with and additional phase factor
${\rm e}^{i\epsilon}$ for everytime they cross the (periodic) boundary from left to right and ${\rm e}^{-i\epsilon}$ if  they cross it
in opposite direction.

The  method of images also permits   to  use unconstrained 
path integrals  to  describe systems with
non-trivial boundary conditions \cite{groschea}.
However, in the case higher dimensions   the method is not useful in the presence of  non symmetric boundaries 
and  the path integral cannot be
defined by a  simple prescription as in the Feynman original formulation. 

For some boundary conditions of one dimensional systems  it  is possible to use another 
method based on the path integral approach with simpler boundary conditions but with an additional singular potential in the Euclidean action
\cite{Posilicano}. 
Two interesting  cases are the following.

v) Quasi-periodic boundary conditions. 

The selfadjoint extension corresponding to the unitary operator
\be U_{\alpha}=\begin{pmatrix} \cos\alpha\, &  \sin \alpha\cr
  \sin\alpha\, & -\cos \alpha
\end{pmatrix}
\label{quasi}
\ee
is defined in the domain of quasiperiodic wave functions
\be
\varphi(1)={\tan \frac\alpha{2}}\,\varphi(0);\qquad {\varphi}'(1)={{\rm cotan} \frac\alpha{2}}{\,}{\varphi}'(0).
\ee

In this case the resolvent, Euclidean time propagators and path integral can be identified with those of 
 periodic boundary conditions
with an extra potential term in the action corresponding to a $\delta'(x)$ singular interaction.

vi) A  circle with  a defect point.


The Hamiltonian with boundary conditions corresponding to the unitary matrix
\begin{equation}
U =\frac{1}{2-ia}
\begin{pmatrix}
{i a }&{2}\cr
{2}&{ia}
\end{pmatrix}
\label{delta}
\end{equation}
can be thought as equivalent to the Hamiltonian  of a delta function $a \delta(x)$ potential acting on a particle moving  a circle with the
usual periodic boundary conditions  $U=\sigma_1$ \cite{agm, fulop} 
The corresponding formulae for the propagator kernel can be
also written in a compact way \cite{fulop, groscheb}.

However, the method is only restricted to similar  cases and  for generic  boundary conditions a closed form 
expression is not available. 
In higher dimensions the number of boundary conditions for
which the path integral method is useful to describe the
quantum evolutions is even more  limited. 

\section{Conclusions}
In summary, it is possible to generalise the   Feynman  approach
to describe the dynamics of quantum systems constrained
to bounded domains.
The boundary itself
has to  be considered from this point of view a genuine quantum
device and  transitions amplitudes are required to implement
the effect of the boundary into the path integral approach.
 For some  boundary conditions 
the modification of the path integral formula includes a phase factor
or a boundary weight for the trajectories which reach the boundary.
However, the method becomes not useful for generic boundary conditions
because the prescription becomes very intricate.
This fact is a consequence of the enhancement of genuine quantum
effects by the presence of the boundary. 

\medskip

\section*{Acknowledgements}

\medskip

We thank M. Aguado,  D. Garc{\'{\i}}a  Alvarez,  G. Marmo and 
P. A. G. Pisani for interesting discussions.  M. A. thanks to E. C. G. Sudarshan
for discussions and inspiration.
This work is partially supported by CICYT (grant FPA2006-2315)
and DGIID-DGA (grant2007-E24/2).

\end{document}